\documentclass[twocolumn,
superscriptaddress,
prl,
showpacs,
preprintnumbers,
amsmath,amssymb]{revtex4}

\usepackage[pdftex]{graphicx}
\usepackage{epsfig}
\usepackage{bm}
\usepackage{subfigure}
\usepackage{color}

\begin{document}

\title{Electric field reduced charging energies and two-electron bound excited states of single donors in silicon}

\author{R. Rahman*}
\affiliation{Advanced Device Technologies, Sandia National Laboratories, Albuquerque, NM 87185, USA}

\author{G. P. Lansbergen}
\affiliation{Kavli Institute of Nanoscience, Delft University of Technology, Lorentzweg 1, 2628 CJ Delft, The Netherlands}

\author{J. Verduijn}
\affiliation{Kavli Institute of Nanoscience, Delft University of Technology, Lorentzweg 1, 2628 CJ Delft, The Netherlands}
\affiliation{Centre for Quantum Computation and Communication Technology, School of Physics, University of New South Wales, Sydney, New South Wales 2052, Australia}

\author{G. C. Tettamanzi}
\affiliation{Kavli Institute of Nanoscience, Delft University of Technology, Lorentzweg 1, 2628 CJ Delft, The Netherlands}
\affiliation{Centre for Quantum Computation and Communication Technology, School of Physics, University of New South Wales, Sydney, New South Wales 2052, Australia}

\author{S. H. Park}
\affiliation{Network for Computational Nanotechnology, Purdue University, West Lafayette, IN 47907, USA}

\author{N. Collaert}
\affiliation{Inter-University Microelectronics Center (IMEC), Kapeldreef 75, 3001 Leuven, Belgium}

\author{S. Biesemans}
\affiliation{Inter-University Microelectronics Center (IMEC), Kapeldreef 75, 3001 Leuven, Belgium}

\author{G. Klimeck}
\affiliation{Network for Computational Nanotechnology, Purdue University, West Lafayette, IN 47907, USA}

\author{L. C. L. Hollenberg}
\affiliation{Centre for Quantum Computation and Communication Technology, School of Physics, University of Melbourne, VIC 3010, Australia}

\author{S. Rogge}
\affiliation{Kavli Institute of Nanoscience, Delft University of Technology, Lorentzweg 1, 2628 CJ Delft, The Netherlands}
\affiliation{Centre for Quantum Computation and Communication Technology, School of Physics, University of New South Wales, Sydney, New South Wales 2052, Australia}

\date{\today}

\begin{abstract}
We present atomistic simulations of the $D^0$ to $D^-$ charging energies of a gated donor in silicon as a function of applied fields and donor depths and find good agreement with experimental measurements. A self-consistent field large-scale tight-binding method is used to compute the $D^-$ binding energies with a domain of over 1.4 million atoms, taking into account the full bandstructure of the host, applied fields, and interfaces. An applied field pulls the loosely bound $D^-$ electron towards the interface and reduces the charging energy significantly below the bulk values. This enables formation of bound excited $D^-$ states in these gated donors, in contrast to bulk donors. A detailed quantitative comparison of the charging energies with transport spectroscopy measurements with multiple samples of arsenic donors in ultra-scaled FinFETs validates the model results and provides physical insights. We also report measured $D^-$ data showing for the first time the presence of bound $D^-$ excited states under applied fields.   

\end{abstract}


\maketitle 

\section{I. Introduction}
A single gated donor atom in silicon has attracted considerable attention over the last decade as a promising quantum information processing unit in solid-state \cite{Kane1}. Among other factors, such donor qubits benefit from exceptionally long spin coherence times \cite{Tryshkin}, compatibility with the complementary metal-oxide-semiconductor (CMOS) technology, and accessibility to controllable atomic physics in the solid-state. Some promising proposals for donor qubits are based on encoding quantum information in the nuclear \cite{Kane1} or electronic spin \cite{Vrijen, Hill, architecture}, or in the molecular charge states of two phosphorus donors \cite{Hollenberg1}. Besides applications in quantum computing, discrete dopants are becoming increasingly important in nanoscale electronics, as they strongly affect the sub-threshold current-voltage characteristics of ultra-scaled MOSFETs \cite{Sanquer3, Asenov}. Other applications of dopants in nanoelectronics are also emerging, such as the proposal for classical logic devices based on resonant tunneling through the donor states \cite{Rogge2, Rogge3}.  

It is well-known that Group V donors in silicon can bind either 1 electron and form a neutral $D^0$ state, or bind 2 electrons and form a negatively charged $D^-$ state. In bulk donors, the $D^-$ ground state is a singlet weakly bound at 1.7-2.0 meV below the conduction band (CB) edge, corresponding to a charging energy (CE) of about 43 meV for a phosphorus (P) impurity and 52 meV for an arsenic (As) impurity \cite{dminus}. Recently, resonant tunneling through the $D^0$ ground and excited states has been observed in different experiments \cite{Rogge, Sanquer3, Dzurak2} with devices fabricated from both top-down \cite{Jamieson} and bottom-up \cite{Schofield} approaches. Most of the measurements of the D- state in gated devices show a significantly reduced charging energy in the range of 25-35 meVs, a conundrum still largely unresolved although some progress has been made using effective mass treatments \cite{Tucker, Hollenberg2, Calderon, Peeters}. In this work, we compute the $D^-$ charging energy of a donor under applied fields using self-consistent field tight-binding (TB) including over 1 million atoms in the spatial simulation domain. We compare the computed CEs with measured data on single dopants in seven FinFET devices by simulating the atomic environment of the devices. Despite being a mean-field technique, this method enables an accurate quantitative description of the charging energies as it captures the details of confinement geometries and valley-orbit interaction from a full bandstructure technique. The method solves the Poisson equation iteratively with the atomistic TB Hamiltonian for charge self-consistency, and represents an advancement over general TB as computational hurdles had to be overcome to solve million atom systems in reasonable time. 

Although no excited $D^-$ states have ever been observed in bulk samples \cite{dminus}, our measurement shows for the first time the presence of bound excited states including a triplet state, which can be ascertained by the phenomena of lifetime enhanced transport \cite{dminus_rogge, eriksson}. This is a consequence of the applied electric field that lowers the charging energy and enables an excited manifold to form below the CB. A two-electron bound triplet state may enable easier ways to perform spin readout through spin blockade between a donor electron and an electron from a local reservoir (2DEG) state. Previous proposals of donor based spin readout have relied on using either Group VI donors \cite{Kane2, Calderon2} or a pair of Group V donors \cite{Hollenberg2}, systems in which two-electron triplet states exist. Our result shows such spin readout may be feasible with the more conventional P/As donor species under strong applied fields. With the recent demonstration of single shot spin readout of a single electron in silicon \cite{Andrea}, this has become even more significant in the context of quantum computing.

This paper is organized as follows. In Section II, we describe the theoretical method for computing the charging energy in detail. We also elaborate on the experimental technique employed to measure the $D^-$ charging energies and excited states in FinFETs. In Section III, we discuss the computed results and how they compare to the measurements. Section IV concludes this work.  

\section{II. Method}

\subsection{A. Theoretical Model}
In the atomistic TB method employed here, the Hamiltonian is represented in a basis of 10 localized atomic orbitals per atom with the sp$^3$d$^5$s* nearest neighbor model \cite{slater, Harrison}. The Hamiltonian parameters of the host material have been optimized using genetic algorithm with analytically derived constraints to fit critical features of the host band structure \cite{Klimeck1, boykin}. Once the TB model parameters of the host are obtained, they are generally transferable to a whole range of device simulations, as benchmarked in a number of earlier works \cite{Klimeck2, Rahman2, kharche}. The full TB Hamiltonian of more than a million silicon atoms is diagonalized using a parallel Lanczos algorithm to obtain any number of lowest lying eigenvalues and wavefunctions. 
   
The TB Hamiltonian of the host and the donor in an applied electric field $\vec{F}$ is given by,

\begin{equation}
H = H_0-V_{\rm D}+e\vec{F}\cdot(\vec{r}-\vec{r_0})+V_{\rm SCF} 
\label{eq1}
\end{equation} 

\noindent where $H_0$ is the TB Hamiltonian of the host material silicon, and $V_{\rm D}$ is the central-cell corrected Coulomb potential energy of the donor. The 3rd term in eq (\ref{eq1}) is that of a constant electric field, whereas the last term $V_{\rm SCF}$ is the potential energy due to electron-electron repulsion. 

Central-cell correction represents the deviation of the donor potential from its $1/r$ form near the donor core \cite{Kohn}, and is responsible for the different $D^0$ binding energies of different Group V donor species \cite{Ramdas}. The central-cell corrected singular-like potential near the donor core produces coupling between the six conduction band (CB) minima Bloch states, and results in a splitting of the orbital ground state of the donor from the excited states. This is commonly known in literature as the `valley-orbit' splitting \cite{VO}. The potential energy of the donor to first order in our model is given by, 

 \begin{equation}
 V_{\rm D}(\vec{r} \neq \vec{r_0})=\frac{e^2}{4 \pi \epsilon_{Si}|\vec{r}-\vec{r_0}|}, V_{\rm D}(\vec{r_0})= U_0
 \label{eq2}
\end{equation} 

\noindent where $\vec{r_0}$ is the location of the donor, $e$ the electronic charge, $\epsilon_{Si}$ the dielectric constant of silicon, and $U_0$ is a cut-off potential representative of the central-cell correction to the first order. A detailed model of impurities in TB can be found in Ref \cite{shaikh}.

$V_{\rm SCF}$ is the potential due to a bulk charge density $n(\vec{r})$, and has to be computed self-consistenly from a reduced Poisson equation as shown below. In the first iteration, $V_{\rm SCF}$ is set to zero to obtain the $D^0$ states of the donor. The solution of $H\psi=E\psi$ yields a set of energies $E=\lbrace E_i \rbrace$ and wavefunctions $\psi=\lbrace \psi_i \rbrace$, where $E_1$ and $\psi_1$ are the ground state energy and wavefunction of a donor respectively. For a bulk P donor at zero field, the above method yields $D^0$ binding energy $E_1$ of 45.6 meV below the CB minimum. For an As donor, it is 54 meV \cite{Ramdas}. 
To obtain the $D^-$ orbital energy, we assume the ground state is filled by exactly one electron, described by an electron density $n(\vec{r})=|\psi_1(\vec{r})|^2$. $V_{\rm SCF}$ is then given by,

 \begin{equation}
 V_{\rm SCF}(\vec{r}) = \int \frac{e^2 n(\vec{r'})}{4 \pi \epsilon_{Si}|\vec{r}-\vec{r'}|} \mathrm{d}\vec{r'} 
 \label{eq3}
\end{equation} 
 
\noindent With the new $V_{\rm SCF}$ eq (\ref{eq1}) is solved for a new set of eigenstates and vectors. The process is repeated until $E_1$ and $n(\vec{r})$ both converge. Since eq (\ref{eq1}) with a converged $V_{\rm SCF}$ represents an impurity in silicon with a bound electron, the converged orbital energy $E_1$ represents the binding energy of the second electron below the CB minimum. This method is also described in detail in Ref \cite{Datta}.
 
This technique makes use of a density functional approximation, as the electron-electron interaction potential is expressed as a function of the electron density. Since the $D^-$ ground state resides in a closed-shell singlet, there is no exchange energy between the two electrons. The repulsive Coulomb energy evaluated self-consistently thus offers a good description of the binding energy. However, the method is still an approximate one, as higher order exchange-correlation corrections are ignored. 
 
An exact way to solve this problem is a full configuration interaction (CI) method, in which the exact 2e Hamiltonian is diagonalized in a basis of Slater Determinants describing anti-symmetric 2e configurations built from a complete set of single particle states \cite{SOR}. However, such a method requires a large number $D^0$ orbitals to represent the spatial extent, symmetry, spin and valley configuration of the $D^-$ states making it both impractical and computationally intractable. In other words, the $D^-$ state is not well-represented with a basis comprising of the low lying $D^0$ states, suggesting the need for a self-consistent method to iteratively improve the basis states. Furthermore, the six-fold valley degeneracy in silicon makes the $D^0$ orbital basis set six times as large. 
Moreover, evaluating a large number of coulomb and exchange integrals with atomistic wavefunctions spanning over a million atoms becomes a computational challenge. On the other hand, atomistic wavefunctions are required for a highly accurate description of the single particle basis functions to capture details of bandstrucuture, geometry and interfaces, all of which are critical for a proper description of impurities in silicon \cite{Rahman1}.   

The mean-field method presented here thus serves as an intermediate theory that combines a highly accurate single particle basis states with a mean-field description of the Coulomb interaction, while ignoring exchange-correlation corrections associated with the 2e system. Although this method cannot capture the exact 2e wavefunctions, it provides a very good description of the $D^-$ binding energies, and the $n(\vec{r})$ associated with the added electron, as we describe in the Section III.    
 
As a benchmark of the method, we computed the $D^-$ binding energy of a bulk P donor in silicon. Photoluminescence experiments have measured this energy to be 2.0 meV below the CB minimum \cite{dminus}, which corresponds to a CE of 43.6 meV. In comparison, the SCF TB method described here yields a binding energy of 3.4 meV for a bulk P donor, corresponding to a CE of 42.2 meV. The difference of 1.4 meV from the experimental value can be regarded as a limitation of the theory due to neglecting exchange-correlation corrections and due to the fact that the TB method only considers point charges on an atomistic zincblende lattice. 

Typical simulation times of the 1e donor states from the TB Hamilton for 1.4 million atoms is about 2-3 hours on 40 processors. The computation of $V_{\rm SCF}$ requires 0.5 hours on 40 processors. Typical $D^-$ energies were found to converge in between 10 to 20 iterations of eq (\ref{eq1}) and eq (\ref{eq3}) \cite{Hoon, Sunhee}.   

Most other works on the $D^-$ donor state have been based on describing the weakly bound 2e wavefunction with variational envelope wavefunction in a single valley picture \cite{Tucker, Hollenberg2, Calderon, Peeters}. One exception is Ref \cite{MonteCarlo}, in which a Quantum Monte Carlo approach was used to compute the bulk $D^-$. Two recent works have treated the effect of screening on the $D^-$ energies in the presence of metallic gates and hetero-interfaces using an effective mass approach \cite{Calderon, Peeters}, and have provided qualitative trends. However, the field and depth dependence of the donors that are directly relevant in measurements need to be investigated in detail from a more quantitative approach. 
 
\subsection{B. Experimental Technique}

\begin{figure}[htbp]
\includegraphics[width=3in]{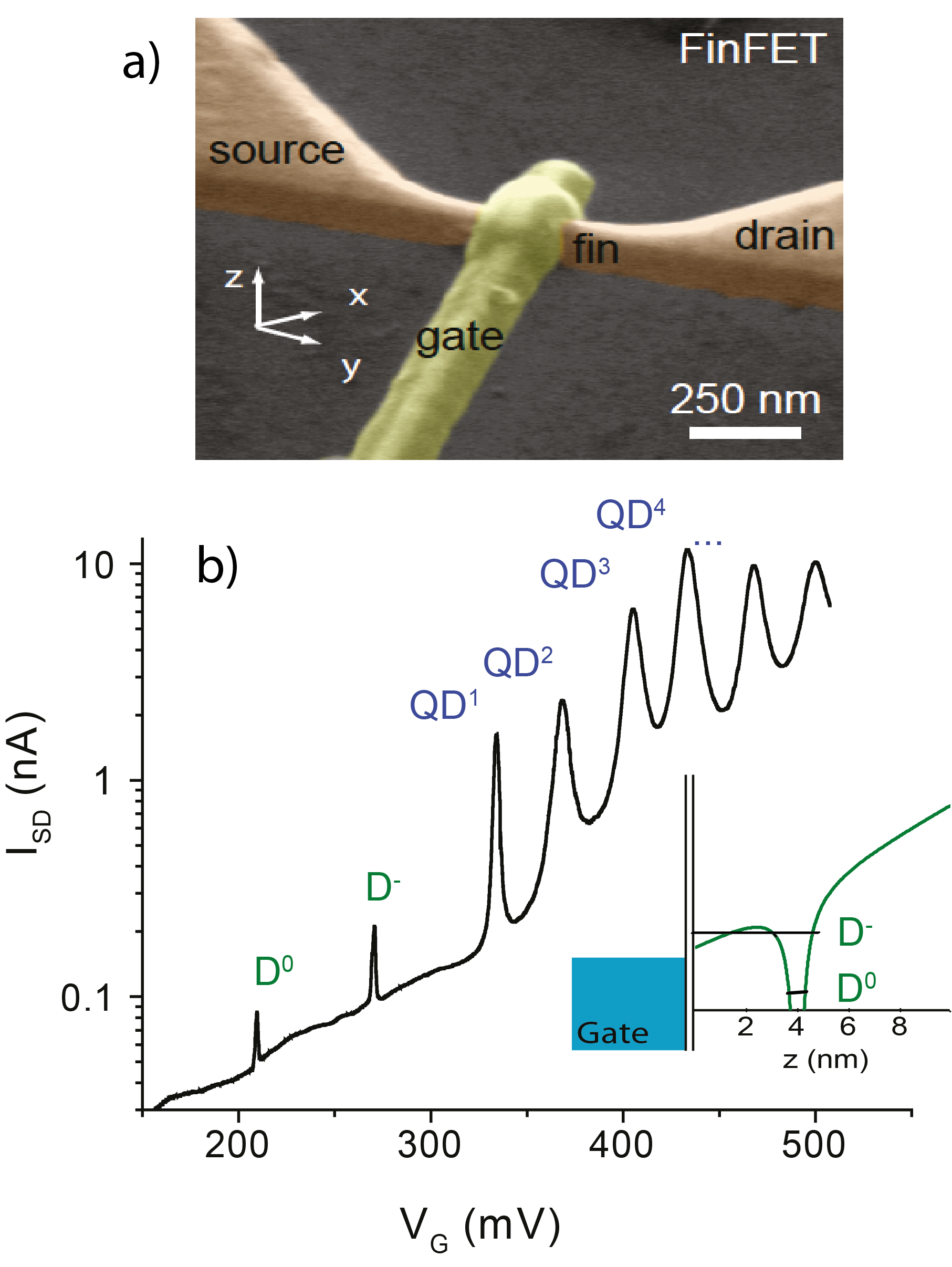}
\caption{
a) Colored SEM image of a silicon FinFET device.
b) Coulomb blocked transport in a FinFET transistor. Two separate charge islands can exist inside the transistor; a quantum dot confined by the triangular potential at the gate interface and residual barriers in the access regions between source/drain  and channel and a donor/well system confined by the donors Coulomb potential and a well at the interface. The localized states formed in these charge islands are denoted by QD$^n$ and D$^0$/D$^-$ respectively. Inset: The potential landscape in a cross-section from the gate to the channel. The gate electric field induces a triangular potential at the interface. An (accidental) donor atom in the channel forms a Coulombic potential on top of the gate potential (green) that can bind two electrons. 
}
\vspace{-0.5cm}
\label{fi1}
\end{figure}


The measurements presented in this work were performed on single As donors in silicon FinFETs, consisting of a silicon nanowire with a gate covering three faces of the body (Fig. 1(a)). A thin nitrided oxide layer separates the gate from the channel. In some devices, a single As donor was found in the channel within 5 nm of the oxide interface. We measured the low-temperature sub-threshold current voltage characteristics of about a 100 devices, and selected 7 devices where the transport characteristics are dominated by a single donor atom in the channel. The other devices either showed no sub-threshold signal or a complicated pattern associated with Coulomb interaction between several donors in the channel. The experimental technique is detailed in our earlier works \cite{Rogge, Sellier, VO}. 

Fig. 1(b) shows the source-drain current of device GLG14 \cite{Rogge} at low bias as a function of gate voltage. At any gate voltage where a localized state in the channel is within the bias window defined by source/drain, it gives a contribution to the transport and a peak in the current occurs. As such, we can perform spectroscopy, with the gate voltage being a measure for the energy of the level ($E = \alpha V_{\rm G}$ where $\alpha$ is the electrostatic coupling between the gate and the level). Based on the aforementioned criteria, we identify the first two resonances as the  D$^0$ and $D^-$ charge states of a single As donor. The resonances indicated by QD$^n$ are due to a localized state which is confined by the gate electric field and two barriers in the access regions between source/drain and the channel \cite{Sanquer1, Sanquer2, SellierAPL}. 

The inset of Fig. 1(b) shows a 1D schematic of the potential of a donor close to the interface at a high E-field. The field transforms the confining potential to a (hybridized) mix between the donors Coulomb potential and a triangular well at the interface. This system is thus essentially a gated donor where the donor-bound electrons are partly pulled toward the Si/SiO$_2$ interface. Due to the proximity of the donor to the interface ($<$ 5 nm), it is possible to apply a high E-field without fully ionizing the donor electron to the interface well. As the E-field is increased, the donor electron hybridizes with interface states, and makes a smooth transition to the interface well, in contrast to a bulk donor, where the ionization process is rather abrupt and occurs at much lower fields \cite{Rahman1}. In our previous work, we combined data from excited state spectroscopy of the $D^0$ state with a large-scale tight-binding analysis, to verify the species of the donor \cite{iedm}, their locations and the E-fields they experienced \cite{Rogge}. In this work, we focus on the $D^-$ charging energies for those six devices, as well as a new device sample, to show the trends in charging energies, as well as to show bound excited 2e states in the spectrum. 
   
\section{III. Results and Discussions}

A donor located close to the oxide-silicon interface \cite{Smit, Martins, Calderon2, Rahman1} has been proposed as an important variant of the Kane qubit architecture \cite{Kane1} based on deeply buried donors. In contrast to bulk donors, it is possible to adiabatically pull the donor electron to the interface, and hence to perform precise quantum control and wavefunction engineering by means of an applied gate bias. Hybrid architectures have also been proposed in which electrons from surface bound quantum dots can be selectively shuttled to nearby donor to preserve their spin coherence for longer time-scales \cite{Calderon3}.

\begin{figure}[htbp]
\includegraphics[width=3in]{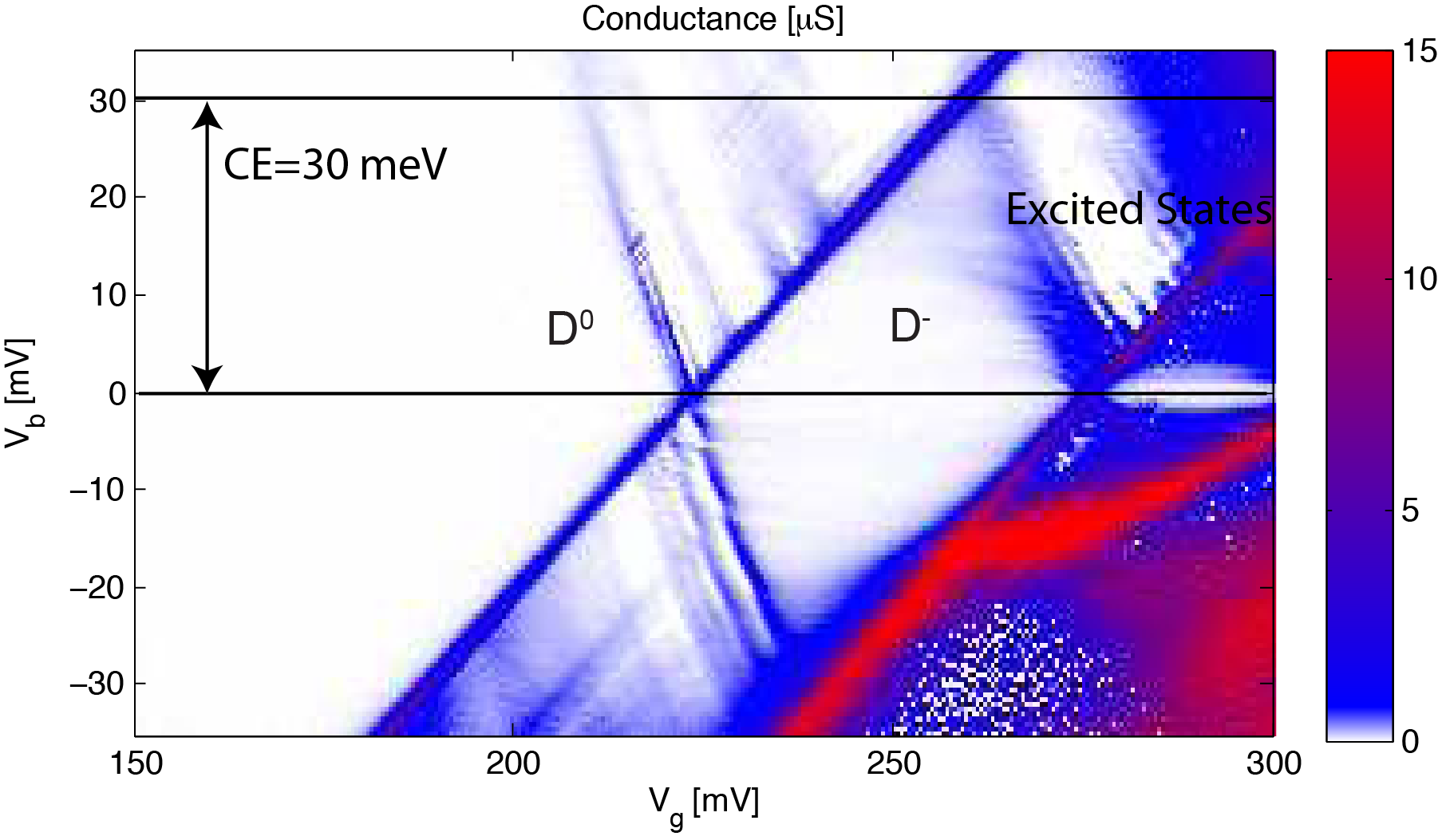}
\caption{
Measured Coulomb diamonds of a device (A18G17) sample showing the $D^0$ and $D^-$ regions. Traces of conductance near the $D^-$ diamond show existence of bound excited states in the two electron spectrum.} 
\vspace{-0.5cm}
\label{fi1}
\end{figure}

In Fig. 2, we show the $D^0$ and $D^-$ regions of the stability diagram of a newly measured device with a gated As donor. Although the CE of a bulk As donor is 52 meV, the measured CE in this sample is 30 meV. We will show that this is a consequence of the applied field that pulls the electron cloud away from each other. Conductance traces through excited states can be observed in both the $D^0$ and $D^-$ regions. While the existence of bound excited states of the $D^0$ are well established \cite{Rogge}, bound $D^-$ excited states are a novel phenomena reported here for the first time. This a consequence of the reduced CE that cause excited manifolds to form below the conduction band. 

Fig. 3(a) shows the 1D potential schematic of the system for various field strengths. As shown in Fig. 1(b), the potential in this system is a superposition of the Coulomb potential of the donor and a linear potential due to the applied E-field. At low field strengths (Fig. 3(a)(i)), the Coulomb potential dominates, and both electrons are bound in the Coulomb well of the donor. This is a bulk-like system with high CE with both electrons confined to a very small region of space. Hence, the electron density around the donor core is high, and the electrostatic repulsion between the electrons stronger, resulting in a higher CE. As the E-field is increased, a triangular well forms at the interface, and the electrons are gradually pulled towards it. At one point, one electron resides in a strongly hybridized state relative to the other, a state which is more delocalized in space, as shown in Fig. 3(a)(ii). The reduced electron densities result in a reduced Coulomb interaction, and the CE decreases. At larger E-fields, both electrons are pulled towards the interface (Fig. 3(a)(iii)), and spreads out more laterally in space.

\begin{figure}[htbp]
\includegraphics[width=3in]{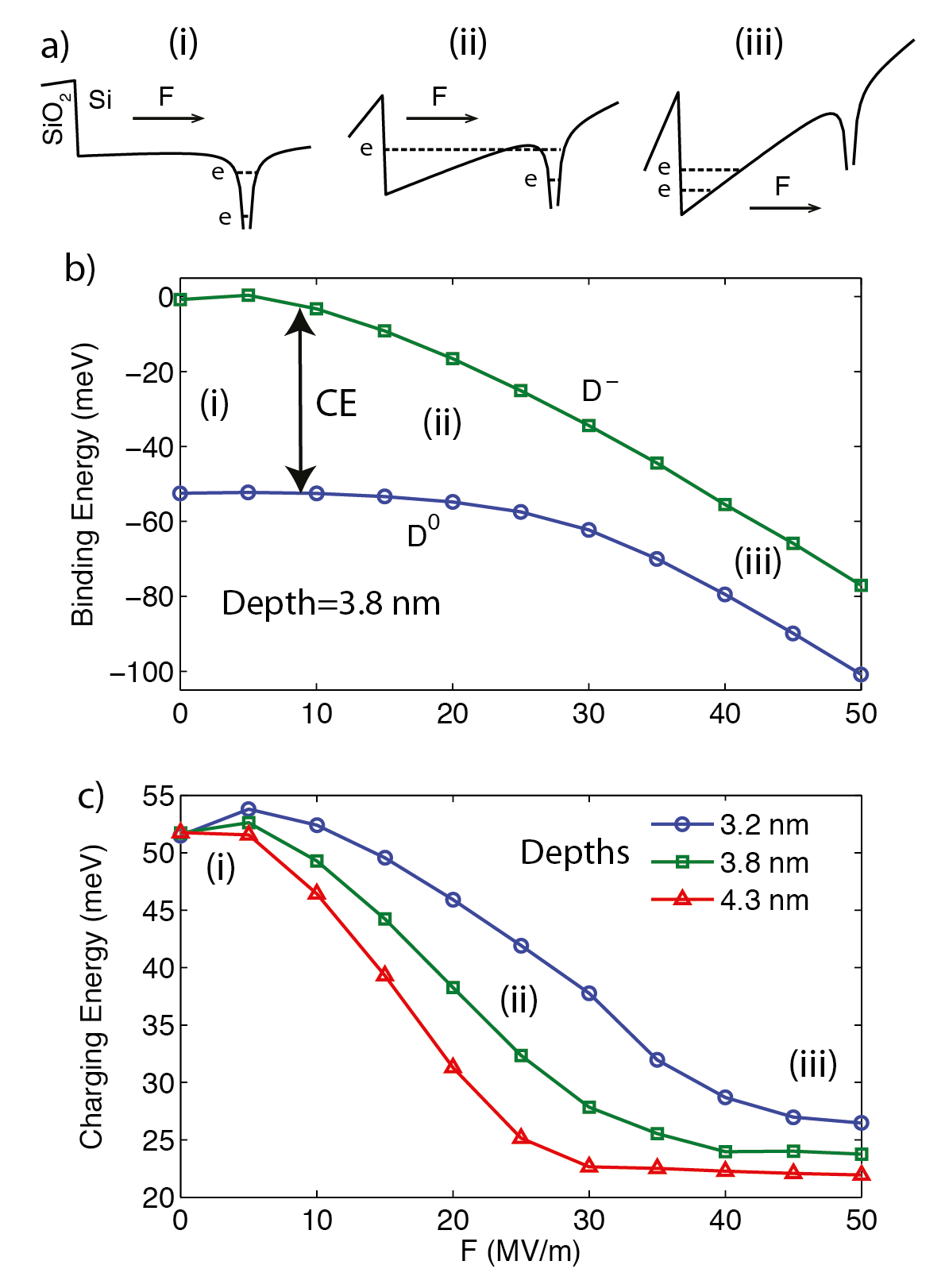}
\caption{(a) 1D schematic of the total potential due to donor and the applied E-field. (i) At low E-field, both the electrons are donor bound. (ii) At increased E-field, one electron is pulled towards the interface, while another remains donor bound. (iii) At higher E-fields, both electrons reside at the interface.
(b) Binding energies of the $D^0$ and $D^-$ states as a function of electric field for an As donor at 3.8 nm depth. The difference between the two energies is the charging energy (CE). (c) CE as a function of electric field for three different donor depths.
}
\vspace{-0.5cm}
\end{figure}

Fig. 3(b) shows the binding energies of $D^0$ and the $D^-$ states as a function of the E-field for an As donor at 3.8 nm from the oxide interface. The energies are expressed relative to the Stark shifted conduction band at the donor site. The difference between the two binding energies is the charging energy, CE. At low fields, corresponding to regime (i) of Fig. 3(a), both electrons are at the donor, and the $D^-$ state is loosely bound below the CB, reminiscent of a bulk-like system. However, the electronic wavefunctions are influenced by the extra confinement due to the nearby interface, and the states are pushed up in energy. As a result, the $D^-$ binding energy in this case is even less than the bulk value of 2 meV below the CB. 

As the E-field is increased, the system gradually moves to the regime (ii) of Fig. 3(a). The electrons begin to hybridize with the interface well states, and both the $D^0$ and $D^-$ binding energies are pushed downwards with applied field, as shown in Fig. 3(b). This orbital Stark effect for the $D^0$ electron has been described in detail in Ref \cite{Rahman1}. In the 2e system, the added electron hybridizes with the interface states sooner than the other electron. As the electron density spreads out more in space, the CE decreases as shown by the decreasing gap between the two energies in Fig 3(b). 

At higher fields, the interface well is occupied by both electrons, as described by regime (iii) of Fig. 3(a). The electronic densities are more laterally extended than before, resulting in a further decrease in CE, which is represented by a constant energy gap between $D^0$ and $D^-$ states, as shown in the region (iii) of Fig. 3(b). However, the electrons are still laterally bounded by the Coulomb potential of the donor, which prevents them from forming a two-dimensional electron gas (2DEG). The lateral confinement potential is stronger for shallower donors. This makes the charge density larger, and the electronic repulsion stronger. Therefore, the CE is expected to increase with decreasing donor depths, as the electrons are pulled into the interface well. In regime (iii), any increase in the vertical field does not influence the electronic wavefunctions significantly, and the CE becomes insensitive to the applied field.    

Fig. 3(c) shows the charging energy as a function of field for three different donor depths. For a specific depth, the CE makes a smooth transition from a bulk-like value of above 50 meV to about 20-30 meV representative of an interfacial confinement regime. Once the electrons are interface bound, the CE becomes field independent. However, due to the lateral Coulomb potential of the donor, the CE values still increase as the donor depth decreases.
In the intermediate field regime, any CE value from about 50 to 30 meV is possible due to hybridization of the donor states with the interface well states \cite{Rahman1}. For smaller donor depths, the transition from the bulk-like CE value to the interface-like CE value is smoother because the tunnel coupling between the two wells is stronger. For a bulk donor, an abrupt step-like transition is expected, as the electrons are ionized without any further confinement. 

Although we have ignored the exact effect of screening in this work, we can qualitatively understand its effect on the $D^-$ binding energies and CEs from Fig. 3(b) and 3(c). Screening essentially modifies the value of the applied E-field due to dielectric mismatch at the silicon-oxide and the gate-oxide interfaces, which results in image charges. If the net screening is dominated by the gate electrode, then the induced image charges would be of opposite polarity to the negatively charged $D^-$ system. As a result, the effective field will be reduced. Since we have used the E-field as a free parameter to investigate the $D^-$ energy over a wide range of field values, the effect of screening can be deduced to first order. For example, if a sample is at 25 MV/m field in Fig. 3(b) and (c), a metallic screening will essentially move the sample to the left of this point in the curve. If the net screening is dominated by the insulating oxide layer, then the induced charges are of the same polarity as the $D^-$ state \cite{Rahman1}. This results in an increased field, which shifts the sample to the right in Fig. 3(b) and 3(c).

However, to obtain an exact quantitative description of screening, the full Poisson equation has to be solved numerically in 3D including the oxide layer and the metal gate, and also self-consistently with the TB Hamiltonian over the whole domain. However, this poses a problem because the atomistic TB Hamiltonian has to be solved over silicon only, as the oxide layer is amorphous in nature and lacks a regular structure. One way to side-step this is to assume a virtual crystal (VC) model of the oxide, and to extend the TB Hamiltonian to include this region, however VC models of SiO$_2$ are still not well-established in literature \cite{Kim, Saraiva}. Although a truncated TB Hamiltonian over the silicon region can be iterated with the Poisson equation over the whole domain, there could be issues relating to charge inconsistencies which would affect the convergence. We have therefore neglected an exact quantitative description of gate screening, and have used the field as the free parameter to investigate its effects approximately. Although some of these limitations can be overcome in principle, this is out of the scope of the present work. 

\begin{figure}[htbp]
\includegraphics[width=2.5in]{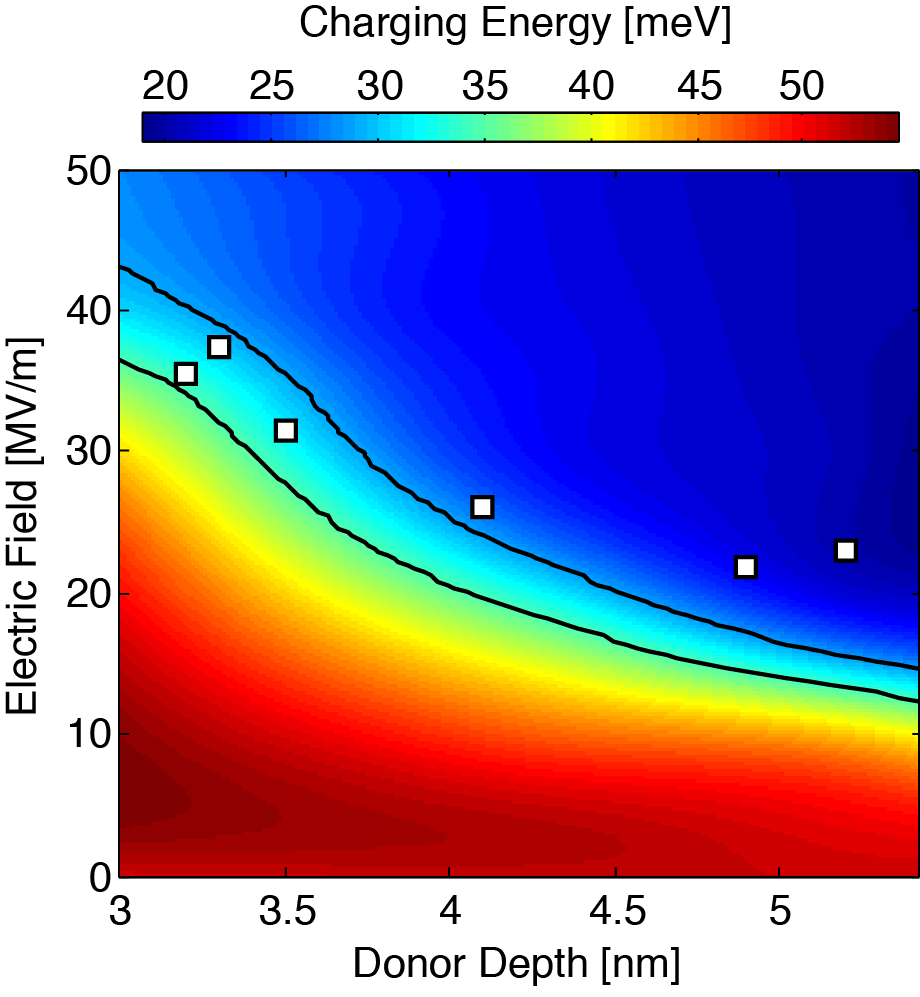}
\caption{Colormap of the modeled charging energy as a function of gate electric field ($F$) and donor depth ($d$). The black traces indicate the region where the modeled charging energy is between 29 and 35 meV, as we also find experimentally. The black data points indicate the positions of the samples in the F-d plane, as determined in a previous publication \cite{Rogge} from their $D^0$ level spectrum. 
}
\vspace{-0.5cm}
\label{fi1}
\end{figure}

In Fig. 4, we show a 2D colormap of the CEs over a range of donor depths (horizontal axis) and E-fields (vertical axis). At high E-fields, the CEs are between 20-25 meV, as indicated by the blue region. This corresponds to interfacial confinement. At low E-fields, the CEs are about 50 meVs, marked by the red region, corresponding to donor bound $D^-$ states. The green and the yellow regions show the intermediate field hybridized regimes. If the donor depth from the interface is large, it takes a smaller field to detune the two wells. Hence, ionization to the interface well takes effect at a lower field value. This is why the blue region of interface-like CEs grow in area from left to right of the plot.

We now compare the measured CEs of six device samples with the computed values. The fields ($F$) and the depths ($d$) of these samples were determined from the $D^0$ excited state transport spectroscopy in our earlier work \cite{Rogge}. We can map these samples on Fig. 4 based on these extracted $F$ and $d$. These data points are marked as black squares. This shows that the $D^-$ energies of these samples were close to the border between the hybridized regime (Fig. 3(a)(ii)) and the interface-bound regime (Fig. 3(a)(iii)), which explains their CEs in the range 29-35 meV. Ideally, all samples should lie between the two black lines plotted in Fig. 4. We attribute the small discrepancy to the fact that we have neglected the exact nature of screening for the $D^-$ state, as discussed above. 
     
\section{IV. Conclusion} 

We have shown through transport spectroscopy measurements on gated As donors in silicon that the charging energies of these donors can be significantly reduced below the bulk values in the presence of an applied E-field. As a consequence, bound excited states are observed for the first time in the $D^-$ spectrum of the donors. This opens up the prospect of performing spin-readout through spin-to-charge conversion between interface and donor bound states using the same Group V donors as the qubits. We present a large-scale self-consistent tight-binding method to compute the charging energies in these nanostructures, taking into account the atomistic details and potentials. The simulations show that the charging energies of the donors are reduced with applied fields, as the electrons hybridize with interface states and delocalize. At low fields, high CEs of above 50 meVs are expected, while at high ionizing fields, the CEs can decrease to about 20-30 meVs. As single donor devices are being fabricated from both top-down and bottom-up approaches, our technique can be used to model the $D^-$ binding energies in a variety of realistic devices.   
 
\begin{acknowledgments}
We acknowledge financial support from the EC FP7 FET-proactive NanoICT projects MOLOC (215750) and AFSiD (214989), the Dutch Fundamenteel Onderzoek der Materie FOM,  the Australian Research Council, the Australian Government, the U.S. National Security Agency (NSA) and the Army Research Office (ARO) under Contract No. W911NF-04-1-0290. This research was conducted by the Australian Research Council Centre of Excellence for Quantum Computation and Communication Technology (Project number CE110001027). NEMO 3D was initially developed at JPL, Caltech under a contract with the NASA. Sandia is a multiprogram laboratory operated by Sandia Corporation, a Lockheed Martin Company, for the United States Department of Energy's National Nuclear Security Administration under Contract No. DE-AC04-94AL85000. NSF funded NCN/nanoHUB.org computational resources were used. 
\end{acknowledgments}

Electronic address: rrahman@sandia.gov


\vspace{-0.5cm}

\begin{thebibliography}{100}  

\bibitem{Kane1} B. E. Kane, Nature, {\bf{393}}, 133 (1998).

\bibitem{Tryshkin} A. M. Tyryshkin, S. A. Lyon, A. V. Astashkin, and A. M. Raitsimring, Phys. Rev. B {\bf{68}}, 193207 (2003). 

\bibitem{Vrijen} Rutger Vrijen, Eli Yablonovitch, Kang Wang, Hong Wen Jiang, Alex Balandin, Vwani Roychowdhury, Tal Mor, and David DiVincenzo, Phys. Rev. A {\bf{62}}, 012306 (2000).

\bibitem{Hill} C. D. Hill, L. C. L. Hollenberg, A. G. Fowler, C. J. Wellard, A. D. Greentree, and H.-S. Goan, Phys. Rev. B {\bf{72}}, 045350 (2005).

\bibitem{architecture} L. C. L. Hollenberg, A. D. Greentree, A. G. Fowler, and C. J. Wellard, Phys. Rev. B {\bf 74}, 045311 (2006).

\bibitem{Hollenberg1} L. C. L. Hollenberg, A. S. Dzurak, C. Wellard, A. R. Hamilton, D. J. Reilly, G. J. Milburn, and R. G. Clark, Phys. Rev. B {\bf{69}}, 113301 (2004).

\bibitem{Sanquer3} M. Pierre, R. Wacquez, X. Jehl, M. Sanquer, M. Vinet, and O. Cueto, Nature Nanotechnology {\bf 5}, 133 (2010).

\bibitem{Asenov} Scott Roy and Asen Asenov, Science Vol. {\bf 309}, 388-390 (2005).

\bibitem{Rogge2} M. Klein, J. A. Mol, J. Verduijn, G. P. Lansbergen, S. Rogge, R. D. Levine, and F. Remacle, Appl. Phys. Lett. {\bf{96}}, 043107 (2010).

\bibitem{Rogge3} Yonghong Yan, J. A. Mol, J. Verduijn, S. Rogge, R. D. Levine, and F. Remacle, J. Phys. Chem. C, {\bf{114}}, 20380 (2010).

\bibitem{dminus} M. Taniguchi and S. Narita, Solid State Communications, Vol. 20, Issue 2, Pages 131-133 (1976).

\bibitem{Rogge} G. P. Lansbergen, R. Rahman, C. J. Wellard, I. Woo, J. Caro, N. Collaert, S. Biesemans, G. Klimeck, L. C. L. Hollenberg, and S. Rogge, Nature Physics {\bf{4}}, 656 (2008).

\bibitem{Dzurak2}Kuan Yen Tan, Kok Wai Chan, Mikko Mottonen, Andrea Morello, Changyi Yang, Jessica van Donkelaar, Andrew Alves, Juha-Matti Pirkkalainen, David N. Jamieson, Robert G. Clark, and Andrew S. Dzurak, Nano Lett. {\bf 10}, 11 (2010).

\bibitem{Jamieson} D. N. Jamieson, C. Yang, T. Hopf, S. M. Hearne, C. I. Pakes, S. Prawer, M. Mitic, E. Gauja, S. E. Andresen, F. E. Hudson, A. S. Dzurak, and R. G. Clark, Appl. Phys. Lett. {\bf{86}}, 202101 (2005).

\bibitem{Schofield} S. R. Schofield, N. J. Curson, M. Y. Simmons, F. J. Rue§, T. Hallam, L. Oberbeck, and R. G. Clark, Phys. Rev. Lett. {\bf{91}}, 136104 (2003).

\bibitem{Tucker} A. Fang, Y. C. Chang, and J. Tucker, Phys. Rev. B {\bf 66}, 155331 (2002).

\bibitem{Hollenberg2} L. C. L. Hollenberg, C. J. Wellard, C. I. Pakes, and A. G. Fowler, Phys. Rev. B {\bf 69}, 233301 (2004).

\bibitem{Calderon} M. J. Calderon, J. Verduijn, G. P. Lansbergen, G. C. Tettamanzi, S. Rogge, and Belita Koiller, Phys. Rev. B {\bf 82}, 075317 (2010).

\bibitem{Peeters} Y. L. Hao, A. P. Djotyan, A. A. Avetisyan, and F. M. Peeters, Phys. Rev. B {\bf 80}, 035329 (2009).


\bibitem{dminus_rogge} G. P. Lansbergen, R. Rahman, J. Verduijn, G.C. Tettamanzi, N. Collaert, S. Biesemans, G. Klimeck, L.C.L. Hollenberg, S. Rogge, arXiv: 1008.1381v1 (2010).

\bibitem{eriksson} Nakul Shaji, C. B. Simmons, Madhu Thalakulam, Levente J. Klein, Hua Qin, H. Luo, D. E. Savage, M. G. Lagally, A. J. Rimberg, R. Joynt, M. Friesen, R. H. Blick, S. N. Coppersmith, and M. A. Eriksson, Nature Physics {\bf 4}, 540 - 544 (2008).

\bibitem{Kane2} B. E. Kane, N. S. McAlpine, A. S. Dzurak, R. G. Clark, G. J. Milburn, He Bi Sun, and Howard Wiseman, Phys. Rev. B {\bf{61}}, 2961 (2000).

\bibitem{Calderon2} M. J. Calderon, Belita Koiller, and S. Das Sarma, Phys. Rev. B {\bf{75}} 161304 (2007).


\bibitem{Andrea} Andrea Morello, Jarryd J. Pla, Floris A. Zwanenburg, Kok W. Chan, Kuan Y. Tan, Hans Huebl, Mikko Mšttšnen, Christopher D. Nugroho, Changyi Yang, Jessica A. van Donkelaar, Andrew D. C. Alves, David N. Jamieson, Christopher C. Escott, Lloyd C. L. Hollenberg, Robert G. Clark, and Andrew S. Dzurak, Nature, {\bf{467}}, 687 (2010).

\bibitem{Harrison} W. A. Harrison, {\it Electronic structure and the properties of solids: the physics of the chemical bond} (Dover Publications Inc. , 1989).

\bibitem{slater} J. C. Slater and G. F. Koster, Phys. Rev. {\bf 94}, 1498 (1954).

\bibitem{Klimeck1} Gerhard Klimeck, Fabiano Oyafuso, Timothy B. Boykin, R. Chris Bowen, and Paul von Allmen, Computer Modeling in Engineering and Science (CMES) Volume 3, No. 5 pp 601-642 (2002), ISSN: 1526-1492.

\bibitem{boykin} B. Boykin, G. Klimeck, and F. Oyafuso, Phys. Rev. B {\bf 69}, 115201 (2004).

\bibitem{Rahman2} Rajib Rahman, Cameron J. Wellard, Forrest R. Bradbury, Marta Prada, Jared H. Cole, Gerhard Klimeck, and Lloyd C. L. Hollenberg, Phys Rev. Lett. {\bf{99}}, 036403 (2007).

\bibitem{Klimeck2} Gerhard Klimeck, Shaikh Ahmed, Hansang Bae, Neerav Kharche, Steve Clark, Benjamin Haley, Sunhee Lee, Maxim Naumov, Hoon Ryu, Faisal Saied, Marta Prada, Marek Korkusinski, Timothy B. Boykin, Rajib Rahman, IEEE Trans. Electron Dev. {\bf{54}}, 2079-2089 (2007).

\bibitem{kharche} N. Kharche, M. Prada, T. B. Boykin, and G. Klimeck, Appl. Phys. Lett. {\bf 90}, 092109 (2007).

\bibitem{Kohn} W. Kohn and J. M. Luttinger, Phys. Rev. {\bf{98}}, 915 (1955).

\bibitem{Ramdas} A. K. Ramdas and S. Rodriguez, Rep. Prog. Phys., Vol. {\bf{44}} (1981).

\bibitem{VO} R. Rahman, J. Verduijn, N. Kharche, G. P. Lansbergen, G. Klimeck, L. C. L. Hollenberg, and S. Rogge, Phys. Rev. B {\bf{83}}, 195323 (2011).

\bibitem{shaikh} S. Ahmed, N. Kharche, R. Rahman, M. Usman, S. Lee, H. Ryu, H. Bae, S. Clark, B. Haley, M. Naumov, F. Saied, M. Korkusinski, R. Kennel, M. McLennan, T. B. Boykin, and G. Klimeck, ÒMultimillion Atom Simulations with NEMO 3-DÓ, Springer Encyclopedia of Complexity and Systems Science, edited by Robert A. Meyers Springer-Verlag GmbH, Heidelberg, 2009, pp. 5745, ISBN: 978-0-387-75888-6.

\bibitem{Datta} S. Datta, {\it Quantum Transport : Atom to Transistor} (Cambridge University Press, 2005).

\bibitem{SOR} Szabo A., and Ostlund, N.S. in {\it Modern Quantum Chemistry-Introduction to Advanced Electronic Structure Theory} (Dover Publications Inc.,1989).

\bibitem{Rahman1} Rajib Rahman, G. P. Lansbergen, Seung H. Park, J. Verduijn, Gerhard Klimeck, S. Rogge, and Lloyd C. L. Hollenberg, Phys. Rev. B, {\bf{80}} 165314 (2009).

\bibitem{Hoon} Hoon Ryu, Sunhee Lee, and Gerhard Klimeck, IEEE proceedings of the 13th International Workshop on Computational Electronics (IWCE), Tsinghua University, Beijing, May 27-29, 2009, pg 1-4, ISBN: 978-1-4244-3925-6. DOI: 10.1109/IWCE.2009.5091082.

\bibitem{Sunhee} Sunhee Lee, Hoon Ryu, Zhengping Jiang, and Gerhard Klimeck, IEEE proceedings of the 13th International Workshop on Computational Electronics (IWCE), Tsinghua University, Beijing, May 27-29, 2009, pg 1-4, ISBN: 978-1-4244-3925-6. DOI: 10.1109/IWCE.2009.5091117.

\bibitem{MonteCarlo} Jun-ichi Inoue, Jun Nakamura, and Akiko Natori, Phys. Rev. B {\bf 77}, 125213 (2008). 

\bibitem{Sellier} H. Sellier, G. P. Lansbergen, J. Caro, and S. Rogge, N. Collaert, I. Ferain, M. Jurczak, and S. Biesemans, Phys Rev. Lett. {\bf{97}}, 206805 (2006).


\bibitem{Sanquer1} F. Boeuf, X. Jehl, M. Sanquer, and T. Skotnicki, IEEE Trans. Nanotechnology {\bf 2}, 144 (2003).

\bibitem{Sanquer2} X. Jehl, M. Sanquer, G. Bertrand, G.Guegan, S. Deleonibus, and D. Fraboulet, IEEE Trans. Nanotechnology {\bf 2}, 308 (2003).

\bibitem{SellierAPL} H. Sellier, G. P. Lansbergen, J. Caro, S. Rogge, N. Collaert, I. Ferain, M. Jurczak, and S. Biesemans, Appl. Phys. Lett. {\bf 90}, 073502 (2007).

\bibitem{iedm} G. P. Lansbergen, R. Rahman, C. J. Wellard, J. Caro, N. Collaert, S. Biesemans, G. Klimeck, L. C. L. Hollenberg, and S. Rogge, ÒTransport-based dopant metrology in advanced FinFETsÓ, IEEE IEDM, San Francisco, December 15Ð17, 2008.

\bibitem{Smit} G. D. J. Smit, S. Rogge, J. Caro, and T. M. Klapwijk, Phys. Rev. B {\bf{70}}, 035206 (2004).

\bibitem{Martins} A. S. Martins, R. B. Capaz, and Belita Koiller, Phys. Rev. B {\bf{69}}, 085320 (2004).


\bibitem{Calderon3} M. J. Calderon, Belita Koiller, Xuedong Hu, and S. Das Sarma, Phys. Rev. Lett. {\bf{96}}, 096802 (2006).

\bibitem{Kim} S. Kim, A. Paul, M. Luisier, T. B. Boykin, G. Klimeck, IEEE Transactions on Electron Devices Vol: PP, Issue {\bf{99}}, 1-10 (2011).

\bibitem{Saraiva} A. L. Saraiva, Belita Koiller, and Mark Friesen, Phys. Rev. B {\bf{82}}, 245314 (2010).

\end{thebibliography}
\end{document}